\documentclass[12pt]{article}
\usepackage[dvips]{graphicx}
\usepackage{amssymb}
\usepackage[hang]{caption2}
\usepackage{psfrag}

\renewcommand{\thefootnote}{\fnsymbol{footnote}}

\def\dspace{\baselineskip = 0.25in}

\newcommand{\inflaton}{inf\mbox{}laton\ }
\newcommand{\inflation}{inf\mbox{}lation\ }
\newcommand{\inflationary}{inf\mbox{}lationary\ }

\begin{document}

\dspace
\begin{titlepage}
\begin{flushright}
BA-03-12\\
\end{flushright}
\vskip 2cm
\begin{center}
{\Large\bf
GUT Scale Inf\mbox{}lation, Non-Thermal Leptogenesis, \\ and Atmospheric Neutrino Oscillations
}
\vskip 1cm
{\normalsize\bf
V. N. Senoguz\footnote{nefer@udel.edu} and
Q. Shaf\mbox{}i\footnote{shaf\mbox{}i@bxclu.bartol.udel.edu}
}
\vskip 0.5cm
{\it Bartol Research Institute, University of Delaware, \\Newark,
DE~~19716,~~USA\\[0.1truecm]}

\end{center}
\vskip .5cm

\begin{abstract}
Leptogenesis scenarios in supersymmetric hybrid \inflation models are considered.
Sufficient lepton asymmetry leading to successful baryogenesis can be obtained if the
reheat temperature $T_r\gtrsim10^6$ GeV and the superpotential coupling parameter $\kappa$
is in the range $10^{-6}\lesssim\kappa\lesssim10^{-2}$.
For this range of $\kappa$ the scalar spectral index $n_s\simeq0.99\pm0.01$.
Constraints from neutrino mixing further restrict the range of $\kappa$ that is allowed.
We analyze in detail the case where the \inflaton predominantly decays into the next-to-lightest 
right handed Majorana neutrino taking into account especially the constraints from atmospheric 
neutrino oscillations.
\end{abstract}
\end{titlepage}

\renewcommand{\thefootnote}{\arabic{footnote}}
\setcounter{footnote}{0}
\setcounter{page}{1}
\section{Introduction}
Supersymmetric hybrid \inflation models \cite{dvaliet.,lazarides} provide a compelling framework
for the understanding of the early universe. They account for the primordial density perturbations 
with a GUT scale symmetry breaking yet without any
dimensionless parameters that are very small.
As in any complete \inflationary
scenario, \inflation in these models should be followed by a succesful reheating
accounting for the observed baryon asymmetry of the universe.

In SUSY hybrid \inflation it is generally preferable (and in many models necessary) to generate the 
baryon asymmetry
via leptogenesis, which is then partially converted into baryon asymmetry by sphaleron effects
\cite{lepto}. 
If the gauge symmetry $G=SO(10)$ or one of its subgroups (where 
inf\mbox{}lation is associated with the breaking of a gauge symmetry $G\to H$),
the \inflaton decays into the right handed neutrinos, whose subsequent out of equilibrium decay
leads to the lepton asymmetry \cite{ls}. The right handed neutrinos could also be produced thermally,
although it is difficult to reconcile the high reheat temperature required by thermal leptogenesis
with the gravitino constraint \cite{gravitino}.

In thermal leptogenesis \cite{thermal} 
the lightest right handed Majorana neutrino $N_1$ 
washes away the previous asymmetry created by the heavier neutrinos. 
If, on the other hand, $N_1$ as well as the heavier neutrinos
are out of equilibrium ($T_r<M_1$), the lepton asymmetry could predominantly result from the 
\inflaton $\chi$ decaying into the next-to-lightest neutrino $N_2$. ($\chi\to N_3\,N_3$
is ruled out by the gravitino constraint.)

In this letter we focus on the latter scenario. It is easier to account for the
observed baryon asymmetry in this case since the asymmetry per right handed neutrino
decay is in general greater than the case where the \inflaton decays into the lightest neutrino,
and unlike thermal leptogenesis there is no washout factor.

The plan of the paper is as follows: In Section 2 we briefly review a class of
supersymmetric hybrid inf\mbox{}lation models. In Section 3 we qualitatively 
discuss leptogenesis scenarios for these models. 
In Section 4 we perform an analysis of the `next-to-lightest' scenario, 
showing numerically that sufficient lepton asymmetry 
can be generated while satisfying, in particular, the constraints from
atmospheric neutrino mixing. 

\section{Supersymmetric Hybrid Inf\mbox{}lation}
In a class of realistic supersymmetric models, 
inf\mbox{}lation is associated with the breaking of either a grand
 unified symmetry or one of its subgroups. Here we will limit ourselves to
supersymmetric hybrid \inflation models \cite{lazarides}.
The simplest such model \cite{dvaliet.} is realized by the renormalizable potential 
(consistent with a $U(1)$ R-symmetry) \cite{copeland}
\begin{equation} \label{super}
W_1=\kappa S(\phi\overline{\phi}-M^{2})
\end{equation}
\noindent where $\phi(\overline{\phi})$ denote a conjugate pair of superfields 
transforming as nontrivial representations of
some gauge group $G$, $S$ is a gauge singlet superfield, and $\kappa$ $(>0)$ is a dimensionless coupling. 
In the absence of supersymmetry breaking, the potential energy minimum 
corresponds to non-zero (and equal in magnitude) vevs $(=M)$
for the scalar components in $\phi$ and $\overline{\phi}$, while the vev of $S$ is zero. 
(We use the same notation for superfields and their scalar components.)
Thus, $G$ is broken to some subgroup $H$.

In order to realize inf\mbox{}lation, the scalar fields $\phi$, $\overline{\phi}$, 
$S$ must be displayed from their present minima.
For $|S|>M$, the $\phi$, $\overline{\phi}$ vevs both vanish 
so that the gauge symmetry is restored, and the tree level
potential energy density $\kappa^{2}M^{4}$ dominates the universe. 
With supersymmetry thus broken, there are radiative
corrections from the $\phi-\overline{\phi}$ supermultiplets 
that provide logarithmic corrections to the potential
which drives inf\mbox{}lation.

The temperature fluctuations $\delta T/T$ turn out to be
 proportional to $(M/M_{P})^2$, where $M$ denotes the symmetry breaking
 scale of $G$, and $M_{P}=1.2\times10^{19}$ GeV is the Planck mass \cite{dvaliet.,lazarides}.
Comparison with the $\delta T/T$ measurements by COBE \cite{cobe} and WMAP \cite{wmap}
shows that the gauge symmetry breaking scale $M$ is naturally of order $10^{16}$ GeV.\footnote{We 
take $(\delta T/T)_h=6.3\times10^{-6}$ \cite{cobe} where h denotes the horizon scale. 
This value corresponds to $A\simeq0.76$ for $n_s=0.99$, in agreement with \cite{wmap}.}

The inf\mbox{}lationary scenario based on the superpotential 
$W_1$ in Eq.\,(\ref{super}) has the characteristic feature that the end of inf\mbox{}lation
essentially coincides with the gauge symmetry breaking. 
Thus, modifications
should be made to $W_1$ if the breaking of $G$ to $H$ leads to the appearance of 
topological defects such as monopoles, strings
or domain walls. For instance, the breaking of $G_{PS}\equiv 
SU(4)_c\times SU(2)_L\times SU(2)_R$ \cite{patisalam} 
to the MSSM by fields belonging to
$\phi(\overline{4},1,2)$, $\overline{\phi}(4,1,2)$ produces magnetic monopoles 
that carry two quanta of Dirac magnetic charge
\cite{Lazarides:1980cc}. As shown in \cite{jeannerot}, one simple resolution 
of the monopole problem is achieved by supplementing $W_1$
with a non-renormalizable term:
\begin{equation} \label{super2}
W_2=\kappa S(\overline{\phi}\phi-\mu^{2})-\beta\frac{S(\overline{\phi}\phi)^{2}}{M^{2}_{S}}\,,
\end{equation}
\noindent where $\mu$ is comparable to the GUT scale, $M_{S}\sim 5\times10^{17}$ GeV 
is a superheavy cutoff
scale, and the dimensionless coefficient $\beta$ is of order unity.  
The presence of the non-renormalizable term enables an inf\mbox{}lationary 
trajectory along which the gauge symmetry is broken. 
Thus, in this `shifted' hybrid \inflation model the magnetic monopoles are inf\mbox{}lated away. 

A variation on these inf\mbox{}lationary scenarios is obtained 
by imposing a $Z_{2}$ symmetry on the superpotential,
so that only even powers of the
combination $\phi\overline{\phi}$ are allowed \cite{smooth}:
\begin{equation} \label{super3}
W_3=S\left(-\mu^2 +\frac{(\phi\overline{\phi})^{2}}{M^{2}_{S}}\right)\,,
\end{equation}
\noindent where the dimensionless parameters $\kappa$ and $\beta$ (see Eq.\,(\ref{super2})) 
are absorbed in $\mu$ and $M_S$.  
The resulting scalar potential possesses two (symmetric) valleys of local minima which are 
suitable for inf\mbox{}lation and along which
the GUT symmetry is broken. The inclination of these valleys is already non-zero
at the classical level and the end of inf\mbox{}lation is smooth, in contrast to inf\mbox{}lation based on the superpotential $W_1$ 
(Eq.\,(\ref{super})). An important consequence is that, as in the case of shifted hybrid inf\mbox{}lation, 
potential problems associated with topological defects are avoided.

In all these models, for the symmetry breaking scale 
$M\sim10^{16}$ GeV, one predicts an essentially scale invariant spectrum 
($0.98\lesssim n_s\lesssim1$ depending on the value of $\kappa$ or of $M$ 
and $|\textrm{d}n_{s}/\textrm{d}\ln k|<10^{-3}$ \cite{ss})
which is consistent with a variety of CMB measurements 
 including the recent WMAP results \cite{wmap,seljak}.

After the end of inf\mbox{}lation, the system falls toward the SUSY vacuum
and performs damped oscillations about it. The inf\mbox{}laton, which we collectively
denote as $\chi$, consists of the two complex scalar fields $(\delta\overline{\phi}+
\delta\phi)/\sqrt{2}$ ($\delta\overline{\phi}=\overline{\phi}-M$, $\delta\phi=\phi-M$)
and $S$, with equal mass $m_{\chi}$. In the presence of $N=1$
supergravity, SUSY breaking is induced by the soft SUSY violating terms in the tree level potential
and $S$ acquires a vev comparable to the gravitino mass $m_{3/2}$ ($\sim$ TeV).
This (mass)$^2$ term provides an extra force driving $S$ to the minimum, but its effect is
negligible for $\kappa\gtrsim10^{-6}$. 

More often than not, SUGRA corrections tend to derail
an otherwise succesful inf\mbox{}lationary scenario by giving rise to 
scalar (mass)$^2$ terms of order $H^2$, where $H$ denotes
the Hubble constant. Remarkably, it turns out that for a canonical SUGRA potential 
(with minimal K\"ahler potential
$|S|^2+|\phi|^2+|\overline{\phi}|^2$), 
the problematic (mass)$^2$ term cancels out for the superpotential $W_1$ in
Eq.\,(\ref{super}) \cite{copeland}. This property also persists 
when non-renormalizable terms that are permitted by the $U(1)_R$ symmetry are included
in the superpotential.\footnote{In general, $K$ is expanded as $K=|S|^2+|\phi|^2+|\overline{\phi}|^2+\alpha|S|^4/M^2_P+\ldots$, 
and only the $|S|^4$ term
in $K$ generates a mass$^2$ for $S$, which would spoil inf\mbox{}lation 
for $\alpha\sim1$ \cite{Panagiotakopoulos:1997ej,Lazarides:1998zf}. 
From the requirement $|S|<M_P$,
one obtains an upper bound on $\alpha\ (\lesssim10^{-3})$ \cite{Asaka:1999jb}.
Since smaller values of $\alpha$ do not effect the dynamics of inf\mbox{}lation significantly
and other terms in $K$ are supressed, we take the K\"ahler potential to be minimal
for simplicity.}

As noted in \cite{linde,ss}, for large values of $\kappa$ the presence of SUGRA corrections 
due to the minimal K\"ahler potential can give
rise to $n_{s}$ values that exceed unity by an amount that is not favored
by the data on smaller scales. SUGRA corrections also become important for tiny
values of $\kappa$.
Nevertheless, they remain ineffective
for a wide range of $\kappa$ ($10^{-6}\lesssim\kappa\lesssim10^{-2}$).
As we shall discuss below, leptogenesis consistent with the observed baryon asymmetry
generally constrains $\kappa$ to a similar range. 

\section{Leptogenesis in SUSY Hybrid Inf\mbox{}lation Models}
The observed baryon asymmetry of the universe can be naturally explained 
via leptogenesis in SUSY hybrid inf\mbox{}lation models.
If inf\mbox{}lation is associated with the breaking of the gauge symmetry $G=SO(10)$ \cite{kyaeet.}
or one of its subgroups 
such as $G_{PS}\equiv SU(4)_{c}\times SU(2)_{L}\times SU(2)_{R}$ \cite{jeannerot} and 
$G_{LR}\equiv SU(3)_c\times SU(2)_{L}\times SU(2)_{R}\times U(1)_{B-L}$ \cite{lss}, the inf\mbox{}laton
decays into right handed neutrino superfields \cite{ls}. 
Their subsequent out of equilibrium decay to lepton and Higgs superfields
leads to the observed baryon asymmetry via sphaleron effects \cite{lepto}.

Before discussing the constraints on $\kappa$ from leptogenesis, we note that an important constraint 
that is independent of the details of the seesaw parameters already arises 
from considering the reheat temperature $T_{r}$ after inf\mbox{}lation,
taking into account the gravitino problem which requires that $T_{r}\lesssim10^{10}$ GeV \cite{gravitino}.
We expect the heaviest right handed neutrino to have a mass of around
$10^{14}$ GeV, which is in the right ball park to provide via the
seesaw a mass scale of about .05 eV to explain the atmospheric neutrino
anomaly through oscillations. Comparing this with \cite{lazarides}
\begin{equation} \label{reheat}
T_r=\left(\frac{45}{2\pi^2 g^*}\right)^{\frac{1}{4}}(\Gamma_\chi\,m_P)^{\frac{1}{2}}\simeq
\frac{1}{16}\frac{(m_P\,m_{\chi})^{\frac{1}{2}}}{M}M_i
\end{equation}
\noindent (where $m_P\simeq2.4\times10^{18}$ GeV is the reduced Planck mass, and the decay rate of the
\inflaton $\Gamma_{\chi}=(1/8\pi)(M^2_i/M^2)m_{\chi}$),
we see that for $m_{\chi}\gtrsim10^{5}$ GeV, $M_i$ 
should not be identified with the heaviest
right handed neutrino, otherwise $T_r$ would be too high \cite{lss}. 
Here we have assumed that the right handed neutrinos $N_i$ acquire mass from a non-renormalizable coupling 
$W\supset(1/m_P)\gamma_i \overline{\phi}\,\overline{\phi}N_i N_i$.\footnote{In this paper
we do not consider the possibility of a renormalizable Yukawa coupling $W\supset y_i \phi N_i N_i$.
This would require $\phi$ to be a $SU(2)_R$ Higgs triplet (or a {\bf 126} of $SO(10)$), 
and the Yukawa couplings would have to
be arranged to yield the intermediate scale 
Majorana masses consistent with the neutrino oscillation parameters.} 
Thus, we require that
\begin{equation} \label{infl}             
\frac{m_{\chi}}{2}\le M_3\le\frac{2M^2}{m_P}
\,.
\end{equation}
\noindent The gravitino constraint expressed by Eq.\,(\ref{infl}) requires $\kappa\lesssim10^{-3}$ 
independent of the details of seesaw parameters for the SUSY hybrid \inflation model 
\cite{Lazarides:1999rt,lazarides,ss}. 
However, in shifted and smooth hybrid \inflation 
the Majorana mass of the heaviest right handed neutrino $M_3\le2M^2/M_S$ can remain 
an order of magnitude greater than the inf\mbox{}laton mass
so that this constraint does not restrict $\kappa$ or $M$ (see Figs.\,5,\,7). 

We now consider the case where the inf\mbox{}laton $\chi$ predominantly 
decays into a right handed neutrino that is heavy compared to the reheat
temperature $T_r$. The ratio of the number density of the right-handed (s)neutrino 
$n_N$ to the entropy density $s$ is given by
\begin{equation} \label{Y_N}
\frac{n_N}{s}\simeq\frac{3}{2}\frac{T_r}{m_{\chi}}B_r\,,
\end{equation}
\noindent where $B_r$ denotes the branching ratio into the right handed neutrino channel. 
The resulting lepton asymmetry is
\begin{equation} \label{Y_L}
\frac{n_L}{s}=\frac{n_N}{s}\epsilon\,,
\end{equation}
\noindent where $\epsilon$ is the lepton asymmetry produced per right 
handed neutrino decay. 

Note that unlike thermal leptogenesis, there is no 
washout factor in non-thermal leptogenesis since lepton number violating 2-body scatterings mediated by 
right hand\-ed neutrinos are out of equilibrium as long as the lightest right handed neutrino mass
$M_1\gg T_r$ \cite{washout}.
More precisely, the washout factor is proportional to $e^{-z}$ where $z=M_1/T_r$ \cite{thermal},
and can be neglected for $z\gtrsim10$.

Suppose that the right handed Majorana masses are hierarchical, with $M_1\ll M_2,M_3$ (but $M_1>T_r$). 
If $M_2,M_3$ are heavier
than $m_{\chi}/2$ the inf\mbox{}laton only decays into $2N_1$. With $B_r=1$, the lepton asymmetry is then
\begin{equation} \label{nls}
\frac{n_L}{s}\simeq\frac{3}{2}\frac{T_r}{m_{\chi}}\epsilon_1\,,
\end{equation}
\noindent with $\epsilon_1$ given by
\begin{eqnarray} \label{epsilon}
 \epsilon_1 
  =
  -
  \frac{1}{8\pi}
  \frac{1}{\left(h h^{\dagger}\right)_{11}}
  \sum_{i = 2,3}
  {\rm Im} 
  \left[
   \{
   \left(
    h h^{\dagger}
    \right)_{1i}
    \}^2
   \right]
   \left[
    f^{V}
    \left(
     \frac{M_i^2}{M_1^2}
     \right)
     +
     f^{S}
     \left(
      \frac{M_i^2}{M_1^2}
      \right)
    \right]
    \,,
\end{eqnarray}
\noindent where
\begin{eqnarray} \label{susy_f}
  f^{V}(x)
  =
  \sqrt{x}\,
  \ln\left(1+\frac{1}{x}\right)
  \,,
  \qquad
  f^{S}(x)
  =
  \frac{2 \sqrt{x}}{x-1}
  \,.
\end{eqnarray}
\noindent Assuming $M_1\ll M_2,M_3$, Eq.\,(\ref{epsilon}) simplifies to
\begin{eqnarray} \label{ep1}
  \epsilon_1 
  &\simeq&
  - \frac{3}{8\pi}
  \frac{1}{\left(h h^{\dagger}\right)_{11}}
  \sum_{i = 2,3}
  {\rm Im} 
  \left[
   \{
   \left(
    h h^{\dagger}
    \right)_{1i}
    \}^2
   \right]
   \frac{M_1}{M_i}
   \,.
\end{eqnarray}
\noindent This formula leads to the upper bound \cite{Hamaguchi:2002vc}
\begin{equation} \label{e1max}
\epsilon_1\lesssim2\times10^{-10}\left(\frac{M_1}{10^6\rm{\ GeV}}\right)\left(\frac{m_{\nu3}}{0.05\rm{\ eV}}\right)\,.
\end{equation}
\noindent 
From the observed baryon to photon ratio $\eta\equiv n_B/n_{\gamma}\simeq6.1\times10^{-10}$ \cite{wmap},
the lepton asymmetry is found to be $|n_L/s|\simeq2.4\times10^{-10}$, where we have used
$n_B/s\simeq\eta/7.04$ \cite{kolb} and $n_L/s=-(79/28)n_B/s$ \cite{khleb}.
Using Eqs.\,(\ref{reheat},\,\ref{nls},\,\ref{e1max}), together with the gravitino constraint 
$T_r\lesssim10^{10}$ GeV, we find that sufficient lepton asymmetry requires
\begin{equation} \label{upper}
m^3_{\chi}\le10^{12}\ {\rm GeV}\left(\frac{m_{\nu3}}{0.05\ {\rm eV}}\right)^2 M^2\,,
\end{equation}
\noindent which yields $m_{\chi}\lesssim10^{15}$ GeV for $M\sim M_{GUT}$.
From $M_1<m_{\chi}/2$, we also obtain the lower bounds
\begin{equation}
T_r\ge1.6\times10^6\ {\rm GeV}\left(\frac{0.05\ {\rm eV}}{m_{\nu3}}\right)\,,
\end{equation}
\begin{equation} \label{lower}
m^3_{\chi}\ge M^2\times 10^{-3}\ {\rm GeV}\left(\frac{0.05\ {\rm eV}}{m_{\nu3}}\right)\,,
\end{equation}
\noindent yielding $T_r\gtrsim10^6$ GeV and $m_{\chi}\gtrsim10^{10}$ GeV or $\kappa\gtrsim10^{-7}$.
This remains so even with degenerate neutrinos, 
since the cosmological bound on the sum of neutrino masses
leads to the limit $m_{\nu i}<0.23$ eV \cite{wmap}.

An alternative scenario \cite{lsv} is the case where $M_1\ll M_2\ll M_3$ (but $M_1>T_r$) and 
$M_2<m_{\chi}/2$. Since the decay width of the inf\mbox{}laton is proportional to $M^2_i$, 
the branching ratios to $2N_1$
and $2N_2$ are $(M_1/M_2)^2$ and $1-(M_1/M_2)^2$ respectively. Thus, provided 
$\epsilon_1\lesssim\epsilon_2$, the contribution
to the lepton asymmetry from $N_1$ is negligible. From Eq.\,(\ref{epsilon}) (with permuted indices) and
Eq.\,(\ref{susy_f})
\begin{eqnarray}
 \epsilon_2 
  \simeq
  -
  \frac{1}{8\pi}
  \frac{1}{\left(h h^{\dagger}\right)_{22}}
  \left[
2\frac{M_1}{M_2}\left(\ln\left[\frac{M_1}{M_2}\right]-1\right)
{\rm Im} 
  \left[
   \{
   \left(
    h h^{\dagger}
    \right)_{21}
    \}^2\right]
  +3\frac{M_2}{M_3}{\rm Im} 
  \left[
   \{
   \left(
    h h^{\dagger}
    \right)_{23}
    \}^2\right]
\right]
\end{eqnarray}
\noindent or, since the first term is negligible for hierarchical Dirac neutrino masses
\begin{eqnarray} \label{ep2}
 \epsilon_2 
  \simeq
  -
  \frac{3}{8\pi}
\frac{M_2}{M_3}
  \frac{{\rm Im} 
  \left[
   \{
   \left(
    h h^{\dagger}
    \right)_{23}
    \}^2\right]}
{\left(h h^{\dagger}\right)_{22}}
\,.
\end{eqnarray}
We can also write this as Eq.\,(\ref{ep1}) with permuted indices 
and recover Eq.\,(\ref{e1max}) with $M_1$
replaced by $M_2$ (see \cite{Hamaguchi:2002vc}, section 2.1).
This indicates that $\epsilon_2$ can easily attain values $\gtrsim\epsilon_1$, 
so that the dominant contribution to the lepton asymmetry is from $N_2$. 
Qualitatively, Eq.\,(\ref{e1max}) 
shows that lepton asymmetry sufficient to meet the observational constraint 
$|n_L/s|\simeq2.4\times10^{-10}$ 
can be generated with reasonable values for the phases.

We conclude this section by summarizing the various constraints on $\kappa$ and
the symmetry breaking scale $M$. As noted in the previous section, for large $\kappa$ (or $M$)
the SUGRA contribution gives rise to $n_{s}$ values that exceed unity by an amount that is not favored
by the data on smaller scales ($n_s\le1$ at $k=0.05$ Mpc$^{-1}$ \cite{wmap}). 
This provides an upper bound on $\kappa$ and $M$ 
for the shifted and smooth hybrid \inflation models \cite{ss}. For SUSY hybrid \inflation with
the renormalizable potential Eq.\,(\ref{super}), the gravitino constraint (Eq.\,(\ref{infl}))
provides a more stringent upper bound. 

For small values of $\kappa$, the SUGRA correction and the soft SUSY breaking
(mass)$^2$ term become important. We find by numerical calculation that 
the primordial density perturbations are too small for $\kappa\lesssim10^{-6}$ for 
SUSY hybrid \inflation and $\kappa\lesssim10^{-7}$ for the shifted model.
Sufficient leptogenesis requires Eqs.\,(\ref{upper},\,\ref{lower}), and these are satisfied
for the range allowed by the constraints above (except for smooth hybrid inf\mbox{}lation,
for which Eq.\,(\ref{upper}) provides a lower bound for $M$). The allowed ranges of $\kappa$
and $M$ are shown in Fig.\,1.\footnote{To be specific we assumed that the SUSY breaking 
induces a mass of 1 TeV for $S$, a mass of 10 TeV would increase the relevant lower bounds by a factor
of $\simeq1.5$.} 

\section{Leptogenesis and \\ Atmospheric Neutrino Oscillations}
Two-family numerical calculations for SUSY hybrid \inflation models discussed here
have been carried out previously in refs.
\cite{lsv,Lazarides:1999rt,jeannerot,smooth3}. Here we update and extend these calculations 
using recent measurements of neutrino oscillation parameters.

A comment is in order whether two-family calculations are physically relevant. We consider the case 
where the $\chi\to N_2\,N_2$ branch is dominant, and
Eq.\,(\ref{ep2}) approximates the lepton asymmetry in terms of two
families only. Since the Dirac masses are assumed to be hierarchical, the $\mu\tau$ block is dominant.
Furthermore, the gauge symmetries suggest a Dirac mixing matrix close to the 
CKM matrix $V_{\rm CKM}$,
which is close to the unit matrix especially in the $\mu\tau$ sector. Under these conditions the 
neutrino mixing matrix
$U_{\rm MNS}$ is approximately obtained by rotating the charged lepton and 
neutral Dirac sectors only in the $\mu\tau$ sector
with respect to the weak basis and diagonalizing the resulting light neutrino mass matrix \cite{lsv}. 

Note that the mixing angle 
obtained this way can only be identified with the atmospheric neutrino mixing angle 
if the mixing angles $\theta_{13}$ and
$\theta_{12}$ are both small. While the solar mixing angle at weak scale is not small \cite{Fukuda:2002pe}, 
its RG evolution can lead to a small angle at the reheat temperature \cite{babu}. 
This occurs for a wide range of CP phases for large $\tan\beta$ ($\simeq m_t/m_b\sim50$
for $G\supset SU(4)_c$) and 
degenerate neutrino masses of $\simeq0.1$ eV
\cite{Antusch:2003kp}. For hierarchical neutrino masses radiative effects on the mixing 
are in general small \cite{Casas:1999tg,Frigerio:2002in}. The solar mixing in this case
could be accounted for by non-diagonal Majorana masses of $\sim10^{-3}$ eV that can arise from higher
dimensional operators \cite{Pati:2002ig}. 

Thus, we can ignore the first family only if we consider the special case of
a small solar mixing angle at large energy scales. For this special case
the lepton asymmetry and the atmospheric neutrino mixing angle can be calculated
without assuming any particular ansatz for the Dirac and Majorana mass matrices.

The lepton asymmetry in this case is given by \cite{lsv}
\begin{equation} \label{ep3}
\frac{n_{L}}{s}=\frac{9\,T_{R}}{16\pi \,m_{\chi}}\,\frac{M_{2}}{M_{3}}\,\frac{%
{\rm c}^{2}{\rm s}^{2}\ \sin 2\delta \ (m_{3}^{D}\,^{2}-m_{2}^{D}\,^{2})^{2}%
}{\langle H_u\rangle(m_{2}^{D}\,^{2}\ {\rm s}^{2}\ +\ m_{3}^{D}\,^{2}{\rm \ c^{2}})}\
\,.
\end{equation}
\noindent Here $\langle H_u\rangle=174\sin\beta~\rm{GeV}$ ($\approx 174~\rm{GeV}$
for large $\tan\beta$), where $\beta=\langle H_u\rangle/\langle H_d\rangle$.
$m_{2,3}^{D}$ are the Dirac masses of the neutrinos (in a basis where they are diagonal
and positive) and ${\rm c}=\cos\theta$,
${\rm s}=\sin\theta$, with $\theta$ and $\delta$ being
the rotation angle and phase which diagonalize the right handed Majorana
mass matrix.  

The light neutrino mass matrix is given by the seesaw
formula:
\begin{equation}
m_{\nu}\approx-\tilde m^{D}\frac{1}{M_{\nu^c}}m^{D},
\label{seesaw}
\end{equation}
\noindent where $m^{D}$ is the Dirac neutrino mass matrix and
$M_{\nu^c}$ the right handed Majorana mass matrix.
The atmospheric neutrino mixing angle $\theta_{23}$
lies \cite{lsv} in the range
\begin{equation}
|\,\varphi -\theta ^{D}|\leq \theta _{23}\leq
\varphi +\theta^{D},\ {\rm {for}\ \varphi +
\theta }^{D}\leq \ \pi /2~,
\end{equation}
\noindent where $\varphi$ is the rotation angle which diagonalizes
the light neutrino mass matrix in the basis where the Dirac
mass matrix is diagonal and $\theta ^{D}$ is the Dirac
mixing angle.

In our analysis we will assume $\theta ^{D}\approx0$ and so take $\varphi=\theta _{23}$.
(From $SU(4)_c$ symmetry, $\theta^D\simeq|V_{cb}|\simeq0.03$ \cite{Lazarides:1990ni}.)
We take $m^D_3$ compatible with $G$, i.e. $m^D_3= m_{\tau}\times \tan\beta$
for $G_{LR}$, and $m^D_3= m_t$ for $SO(10)$ or $G_{PS}$.
These relations hold at $M_{\rm GUT}$, while the relevant values of the parameters
are those at the leptogenesis scale. We estimate the Dirac masses by using the above relations
as approximations, with the values for the quark and lepton masses 
at $T_r=10^9$ GeV given in \cite{Fusaoka:1998vc}.

The light neutrino masses are assumed to be either 
hierarchical with $m_{\nu2}=8.5\times10^{-3}$ eV 
and $m_{\nu3}=0.06$ eV \cite{Fukuda:2002pe,Ahn:2002up,Fogli:2003th}, or degenerate with
$m_{\nu2}=0.104$ eV and $m_{\nu3}=0.122$ eV. Note that the RG evolution of the masses is 
particularly important for the degenerate case. The latter values are calculated 
with $\tan\beta=50$ \cite{Antusch:2003kp}.

Using these Dirac and light neutrino masses, we have numerically calculated the range of 
$\kappa$, the symmetry breaking scale $M$ and the reheat temperature $T_r$ consistent with the observed
baryon asymmetry $n_B/s\simeq8.7\times10^{-11}$ \cite{wmap} and the near maximal atmospheric 
mixing $\sin^{2}2\theta_{23}\gtrsim0.95$ \cite{Fukuda:1998mi,Ahn:2002up,Fogli:2003th}.\footnote{For details of the calculation, we refer the reader to refs. \cite{lsv,Lazarides:1999rt}. Note
that in ref. \cite{lsv}, the small angle MSW solution for $m_2$ was assumed and the atmospheric mixing angle was found to be small for a particular value of $\kappa$. Our results are different but not contradictory, since they hold for different values of $m_2$ and $\kappa$. Also, instead of fixing the heavy Majorana masses and calculating the Dirac masses, we have fixed the Dirac masses and calculated the heavy Majorana masses.} 
For the allowed range of $\kappa$ we also required that $M_2\le m_{\chi}/2<M_3$ and $\gamma_3\le1$ 
where $M_3=2\gamma_3 M^2 / m_P$
($2\gamma_3 M^2 / M_S$ for shifted and smooth hybrid inf\mbox{}lation).
The results are summarized below.
\par \bigskip 
\noindent 1. SUSY hybrid \inflation with 
$G_{LR}\equiv SU(3)_c\times SU(2)_{L}\times SU(2)_{R}\times U(1)_{B-L}$
\par \smallskip
\noindent a. Hierarchical neutrinos: The charged lepton masses at $10^9$ GeV are 
$m_{\mu}=86$ MeV and $m_{\tau}=1.47$ GeV. We set, as an approximation, $m^D_i=m_{i}\times \tan\beta$ with
$\tan\beta=10$. We obtain solutions for $\kappa\sim10^{-3.5}$ with
$T_r\simeq10^9$ GeV, $M_2\simeq10^{10.5}$ GeV and $M_3\simeq10^{13}$ GeV (Fig.\,2).
\par \smallskip
\noindent b. Degenerate neutrinos: With $\tan\beta=50$, solutions are obtained
for $\kappa\sim10^{-3}$ with
$T_r\simeq10^{10.5}$ GeV, $M_2\simeq10^{12}$ GeV and $M_3\simeq10^{13}$ GeV (Fig.\,3).
\par \smallskip
\noindent 2. SUSY hybrid \inflation with $G=SO(10)$
\par \smallskip
\noindent a. Hierarchical neutrinos: We set, as an approximation, $m^D_i=m^q_i$ where 
$m^q_i$ are the quark masses
at $10^9$ GeV ($m_c=0.427$ GeV and $m_t=149$ GeV). We find that there are no solutions
consistent with near maximal atmospheric mixing.
\par \smallskip
\noindent b. Degenerate neutrinos: Solutions are obtained
for $\kappa\sim10^{-4}$ with
$T_r\simeq10^{9}$ GeV, $M_2\simeq10^{11}$ GeV and $M_3\simeq10^{13}$ GeV (Fig.\,4).
\par \bigskip
\noindent 3. Shifted hybrid \inflation with $G_{PS}\equiv SU(4)_c\times SU(2)_L\times SU(2)_R$
\par \smallskip
\noindent a. Hierarchical neutrinos: As in case 2a, there are no solutions with $m^D_i=m^q_i$,
although solutions are obtained with higher values of $m^D_2$. As a numerical example,
$m^D_2=2$ GeV allows solutions (with the coefficient of 
the non-renormalizable coupling $\beta=0.5$) for $\kappa\sim10^{-2.5}$, with the symmetry
breaking scale $M\sim M_{GUT}$ and $T_r\sim10^{9.5}$ GeV (Fig.\,5).
\par \smallskip
\noindent b. Degenerate neutrinos: Solutions for $\kappa\sim10^{-3}$ and $T_r\sim10^9$ GeV are obtained
with $m^D_2=0.427$ GeV and $m^D_3=149$ GeV (Fig.\,6). We have taken $\beta=0.5$, for which
$\kappa\ge3\times10^{-4}$ is required for the \inflationary trajectory.
\par \bigskip 
\noindent 4. Smooth hybrid \inflation with $G_{PS}$ 
\par \smallskip 
\noindent a. Hierarchical neutrinos: As in case 3a, we take $m^D_2=2$ GeV to allow solutions.
The baryogenesis and neutrino
mixing constraints can be satisfied with $T_r\sim10^{10}$ GeV, 
and the heavy Majorana masses are $M_2\sim10^{11}$ GeV and $M_3\sim10^{15}$ GeV (Fig.\,7).
\par \smallskip
\noindent b. Degenerate neutrinos: Taking $m^D_2=2$ GeV to allow solutions,
 the baryogenesis and neutrino mixing constraints can only be satisfied for a narrow range of masses,
with the symmetry breaking scale $M\lesssim10^{16}$ GeV and $T_r\gtrsim10^{10}$ GeV.
\par \bigskip
Note that in our calculations we have assumed $M_1\gg T_r$
so that washout effects are negligible.
Since $M_2/T_r$ turns out to be in the range $10-100$, this assumption conflicts with a strong hierarchy between $M_1$ and $M_2$. Eqs.\,(\ref{ep2}) and (\ref{ep3}) have to be suitably modified for $M_2\sim M_1$. However, the resulting lepton asymmetry does not change significantly, unless the right handed neutrinos are quasi degenerate ($(M_2-M_1)\ll M_1$). On the other hand, if we drop the assumption $M_1\gg T_r$, the constraints to generate sufficient lepton asymmetry become more stringent \cite{thermal,Akhmedov:2003dg}.

\section{Conclusion}
We have reviewed non-thermal leptogenesis in SUSY hybrid \inflation models. 
For the simplest SUSY hybrid \inflation model, sufficient lepton asymmetry 
can be generated provided that the dimensionless coupling constant appearing 
in the superpotential Eq.\,(\ref{super}) satisfies $10^{-6}\lesssim\kappa\lesssim10^{-2}$. 
SUGRA correction to the potential is negligible for this range and the power spectrum is 
essentially scale invariant. For shifted and smooth hybrid inf\mbox{}lation, leptogenesis with larger values
of the coupling constant and the symmetry breaking scale is also possible. 

Constraints from neutrino mixing could further restrict the range of $\kappa$ that is allowed. We
have applied the constraint of maximal (or near maximal) atmospheric mixing, as observed
by Super-Kamiokande and K2K, to the case where the \inflaton predominantly decays into 
the next-to-lightest right handed Majorana neutrino.
We have numerically shown, for this case, that sufficient lepton asymmetry
can still be generated with hierarchical Dirac neutrino masses imposed by the gauge symmetries.

We conclude that SUSY hybrid \inflation models can satisfactorily meet the gravitino and
baryogenesis constraints, consistent with the observed neutrino (mass)$^2$ differences and
near maximal atmospheric neutrino mixing.

\vspace{0.5cm}
\noindent {\bf Acknowledgments}
\\ \noindent
This work was supported by DOE under contract number DE-FG02-91ER40626. \\ Q. S. thanks
George Lazarides for discussions.

\vspace{1cm}
\begin{figure}[htb]
\includegraphics[angle=0, width=12cm]{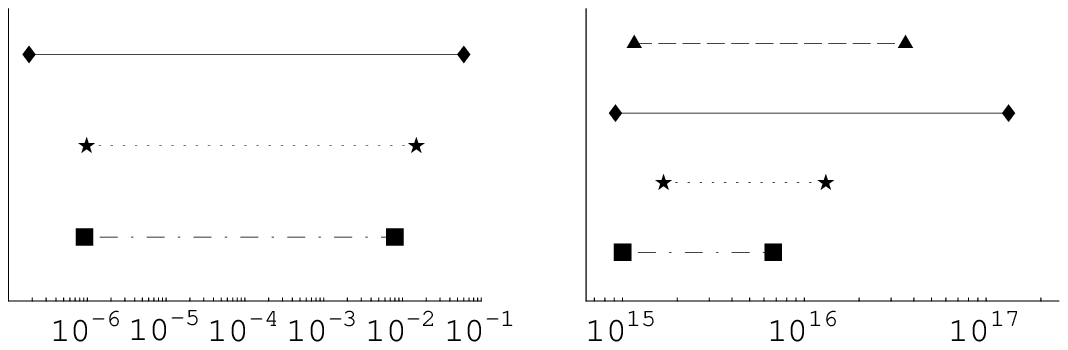}
\caption{\sf The allowed range of the dimensionless superpotential coupling $\kappa$ (left)
and the symmetry breaking scale $M$ (right) for SUSY hybrid \inflation with $G_{LR}$ (dash-dotted line),
SUSY hybrid \inflation with $SO(10)$ (dotted line), 
shifted hybrid \inflation (solid line) and smooth hybrid inf\mbox{}lation (dashed line).}
\end{figure}

\begin{figure}[htb]
\includegraphics[angle=0, width=12cm]{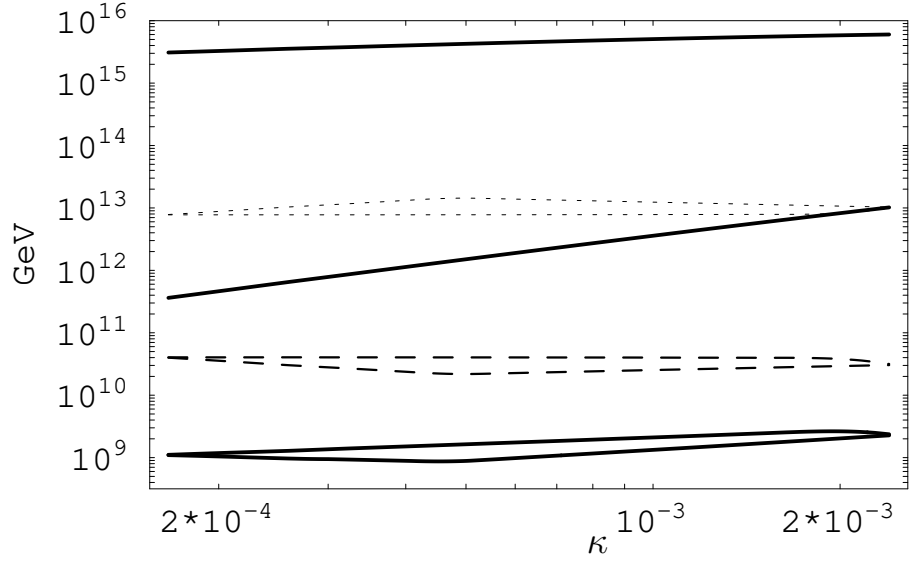}
\vspace{-.8cm}
\begin{center}
{\large \quad $\kappa$}
\end{center}
\caption{\sf From bottom to top, $T_r$, $M_2$ (dashed lines), $m_{\chi}/2$, $M_3$ (dotted lines) 
and $M$ as functions
of $\kappa$, for SUSY hybrid \inflation with $G_{LR}$ and hierarchical left handed Majorana neutrinos. 
The regions for $T_r$, $M_2$ and $M_3$ are bound by the baryon asymmetry
and near maximal atmospheric mixing ($\sin^{2}2\theta_{23}\ge0.95$) constraints.}
\end{figure}

\begin{figure}[htb]
\includegraphics[angle=0, width=12cm]{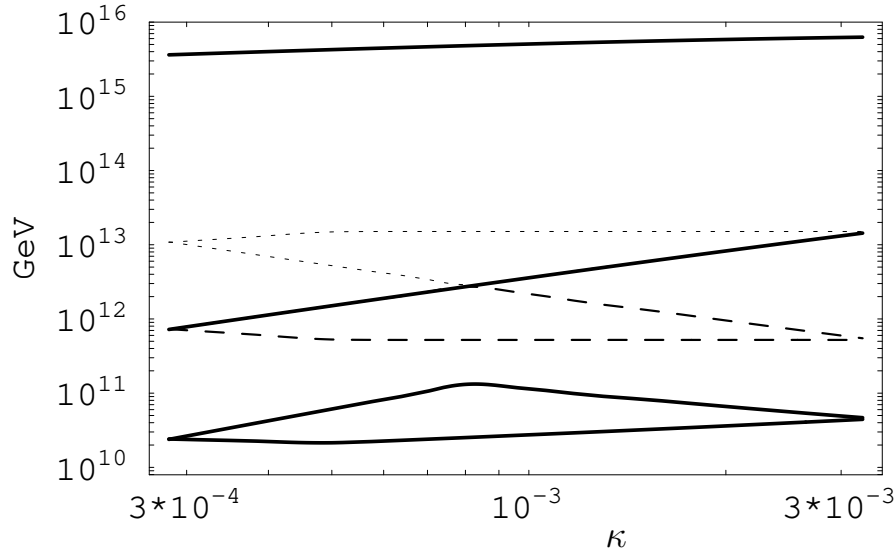}
\vspace{-.8cm}
\begin{center}
{\large \qquad $\kappa$}
\end{center}
\caption{\sf Same as Fig.\,2, for degenerate left handed Majorana neutrinos.
Note that the allowed regions for $M_2$ and $M_3$ are also constrained by
$M_2\le M_{\chi}/2<M_3$.}
\end{figure}

\begin{figure}[htb]
\includegraphics[angle=0, width=12cm]{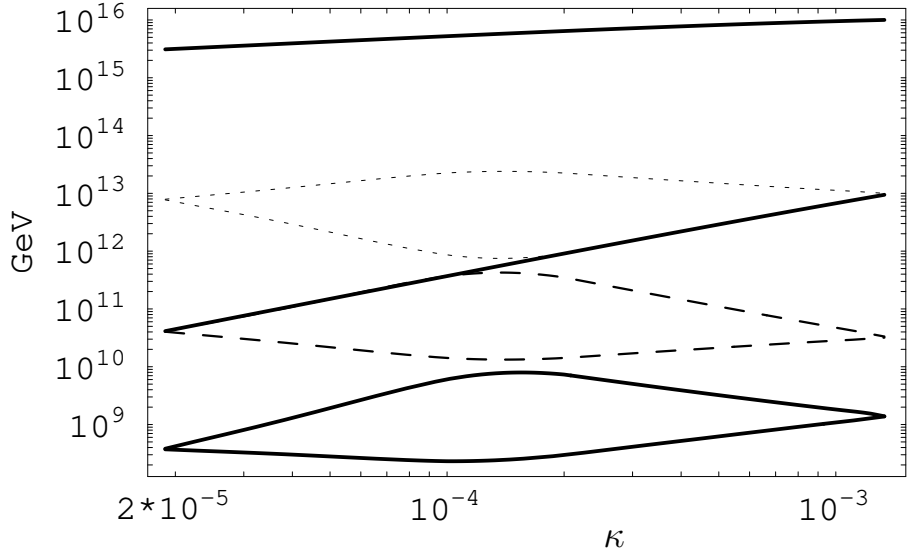}
\vspace{-.8cm}
\begin{center}
{\large \qquad $\kappa$}
\end{center}
\caption{\sf Same as Fig.\,2, for SUSY hybrid \inflation with $G=SO(10)$ and degenerate 
left handed Majorana neutrinos.}
\end{figure}

\begin{figure}[htb]
\includegraphics[angle=0, width=12cm]{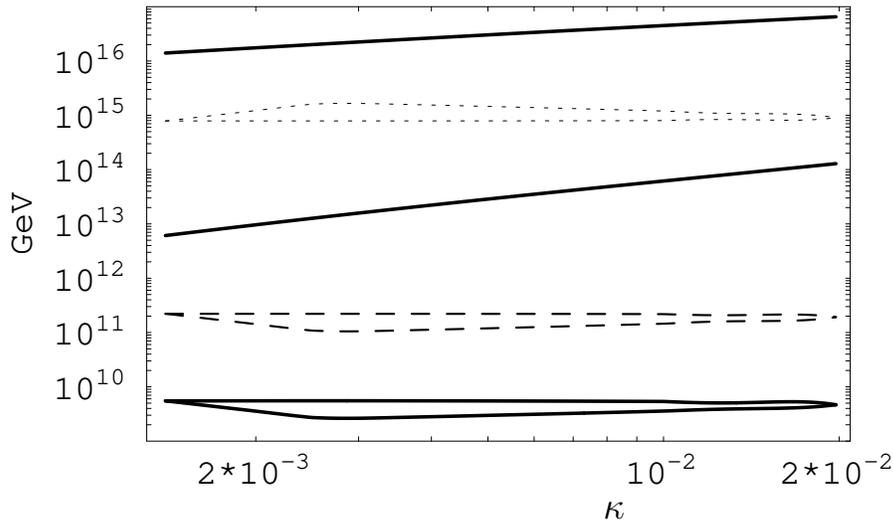}
\vspace{-1cm}
\begin{center}
{\large \qquad $\kappa$}
\end{center}
\caption{\sf Same as Fig.\,2, for shifted hybrid \inflation with $G=G_{PS}$ and hierarchical 
left handed Majorana neutrinos.}
\end{figure}

\begin{figure}[htb]
\includegraphics[angle=0, width=12cm]{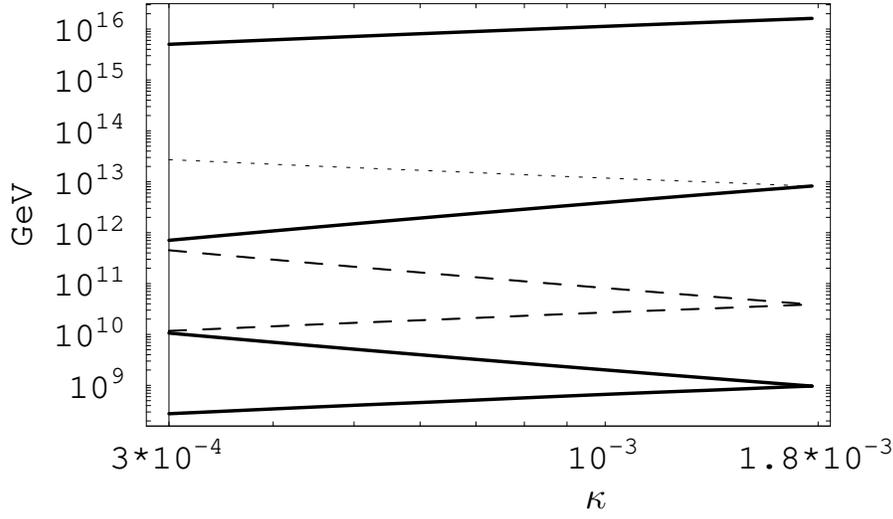}
\vspace{-1cm}
\begin{center}
{\large \quad $\kappa$}
\end{center}
\caption{\sf Same as Fig.\,2, for shifted hybrid \inflation with $G=G_{PS}$ and degenerate 
left handed Majorana neutrinos.
Note that $M_3$ is bound below by $m_{\chi}/2$, and $\kappa>3\times10^{-4}$
is required for the \inflationary trajectory for $\beta=0.5$.}
\end{figure}

\begin{figure}[htb]
\includegraphics[angle=0, width=12cm]{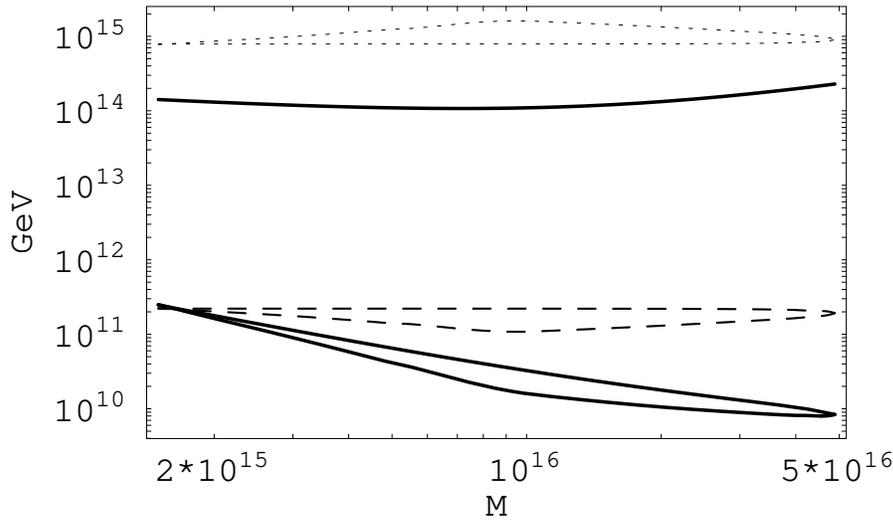}
\caption{\sf From bottom to top, $T_r$, $M_2$ (dashed lines), $m_{\chi}/2$, and $M_3$ (dotted lines) 
as functions of the symmetry breaking scale $M=(\mu M_S)^{1/2}$, for smooth hybrid \inflation 
with $G_{PS}$ and hierarchical left handed Majorana neutrinos.}
\end{figure}
\end{document}